\documentclass[aps,preprint,superscriptaddress,floatfix,footinbib,11pt]{revtex4}
\usepackage{graphicx,epsfig,amsmath,amsfonts,color}
\usepackage[squaren,thinspace,textstyle]{SIunits}
\usepackage{ulem}

\newcommand{\SI}[2]{#1\,#2}
\newcommand{\nm}{\ensuremath{\text{nm}}}
\newcommand{\uV}{\ensuremath{\mu{\text{V}}}}
\newcommand{\dif}{\text{d}}
\newcommand{\ueV}{\ensuremath{\mu{\text{eV}}}}
\newcommand{\mV}{\ensuremath{\text{mV}}}

\newcommand{\V}{\ensuremath{\text{V}}}

\newcommand{\vqpc}{V_{\rm QPC}}
\newcommand{\VQPC}{V_{\rm QPC}}

\renewcommand{\vr}{\ensuremath{V_\text{R}}}
\newcommand{\vl}{\ensuremath{V_\text{L}}}

\newcommand{\algl}{\ensuremath{\alpha_\text{gL}^\text{L}}}
\newcommand{\algr}{\ensuremath{\alpha_\text{gR}^\text{L}}}
\newcommand{\argl}{\ensuremath{\alpha_\text{gL}^\text{R}}}
\newcommand{\argr}{\ensuremath{\alpha_\text{gR}^\text{R}}}

\renewcommand{\emph}[1]{{\it{#1}}}

\definecolor{grey}{rgb}{.6,.6,.6}
\definecolor{darkyellow}{rgb}{.6,.5,0}
\definecolor{darkgreen}{rgb}{0,.6,0}
\definecolor{darkred}{rgb}{.6,0,0}

\newcommand{\aflNRC}{Institute for Microstructural Sciences, National~Research~Council~Canada, Ottawa, ON Canada K1A 0R6}
\newcommand{\aflMunich}{Center for NanoScience and Fakult\"at f\"ur Physik, Ludwig-Maximilians-Universit\"at, Geschwister-Scholl-Platz 1, 80539 M\"unchen, Germany}
\newcommand{\aflMcGill}{Department of Physics, McGill University, Montreal, QC Canada H3A 2T8}
\newcommand{\aflRgb}{Institut f\"ur Experimentelle Physik, Universit\"at Regensburg, D-93040 Regensburg, Germany}

\def\bra#1{\langle#1|}
\def\ket#1{|#1\rangle}

\def\ep{\varepsilon}

\definecolor{DarkRed}{rgb}{0.45,0.08,0}
\definecolor{DarkBlue}{rgb}{0,0.08,0.45}

\def\TT{\mathcal{T}}
\def\ua{\uparrow}
\def\da{\downarrow}

\def\vr{\vec{r}}

\def\vq{\vec{q}}

\renewcommand{\emph}[1]{{\it{#1}}}

\begin{document}

\title{Quantum interference and phonon-mediated back-action \\ in lateral quantum dot circuits}

\author{G.\ Granger}
	\thanks{These authors contributed equally to this work.}
	\affiliation{\aflNRC}
\author{D.\ Taubert}
	\thanks{These authors contributed equally to this work.}
	\affiliation{\aflMunich}
\author{C.\ E.\ Young}
	\affiliation{\aflMcGill}
\author{L.\ Gaudreau}
	\affiliation{\aflNRC}
	\affiliation{D\'epartement de physique, Universit\'e de Sherbrooke, Sherbrooke, QC Canada J1K 2R1}
\author{A.\ Kam}
	\affiliation{\aflNRC}
\author{S.~A.~Studenikin}
	\affiliation{\aflNRC}
\author{P.\ Zawadzki}
	\affiliation{\aflNRC}
\author{D.\ Harbusch}
	\affiliation{\aflMunich}
\author{D.\ Schuh}
\affiliation{\aflRgb}
\author{W.\ Wegscheider}
\affiliation{\aflRgb}
\affiliation{Laboratory for Solid State Physics, ETH Z\"urich, CH-8093 Z\"urich, Switzerland}
\author{Z.\ R.\ Wasilewski}
	\affiliation{\aflNRC}
\makeatletter
\author{A.\ A.\ Clerk}
 	\thanks{The experiments have been supervised collaboratively by A.\ S.\ S.\ and S.\ L.; the theoretical modelling was supervised by A.\ A.\ C. \\ electronic address: Stefan.Ludwig@physik.uni-muenchen.de.}
	\affiliation{\aflMcGill}
\author{S.\ Ludwig}
 	\thanks{The experiments have been supervised collaboratively by A.\ S.\ S.\ and S.\ L.; the theoretical modelling was supervised by A.\ A.\ C. \\ electronic address: Stefan.Ludwig@physik.uni-muenchen.de.}
	\affiliation{\aflMunich}
\author{A.\ S.\ Sachrajda}
 	\thanks{The experiments have been supervised collaboratively by A.\ S.\ S.\ and S.\ L.; the theoretical modelling was supervised by A.\ A.\ C. \\ electronic address: Stefan.Ludwig@physik.uni-muenchen.de.}
	\affiliation{\aflNRC}
\makeatother

\date{\today}


\maketitle

\textbf{Spin qubits have been successfully realized in electrostatically defined,  lateral few-electron quantum dot circuits \cite{Hanson2007,Fujisawa2006,Petta2010,Gaudreau2011}. Qubit readout typically involves spin to charge information conversion, followed by a charge measurement made using a nearby biased quantum point contact \cite{Hanson2007,Field1996,Elzermann2003}. It is critical to understand the back-action disturbances resulting from such a measurement approach \cite{Aguado2000,Young2010}. Previous studies have indicated that quantum point contact detectors emit phonons which are then absorbed by nearby qubits \cite{Khrapai2006,Taubert2008,Gasser2009,Schinner2009,Harbusch2010}.  We report here the observation of a pronounced back-action effect in multiple dot circuits where the absorption of detector-generated phonons is strongly modified by a quantum interference effect, and show that the phenomenon is well described by a theory incorporating both the quantum point contact and coherent phonon absorption. Our combined experimental and theoretical results suggest strategies to suppress back-action during the qubit readout procedure.}

\begin{figure}
\includegraphics[width=0.4\columnwidth]{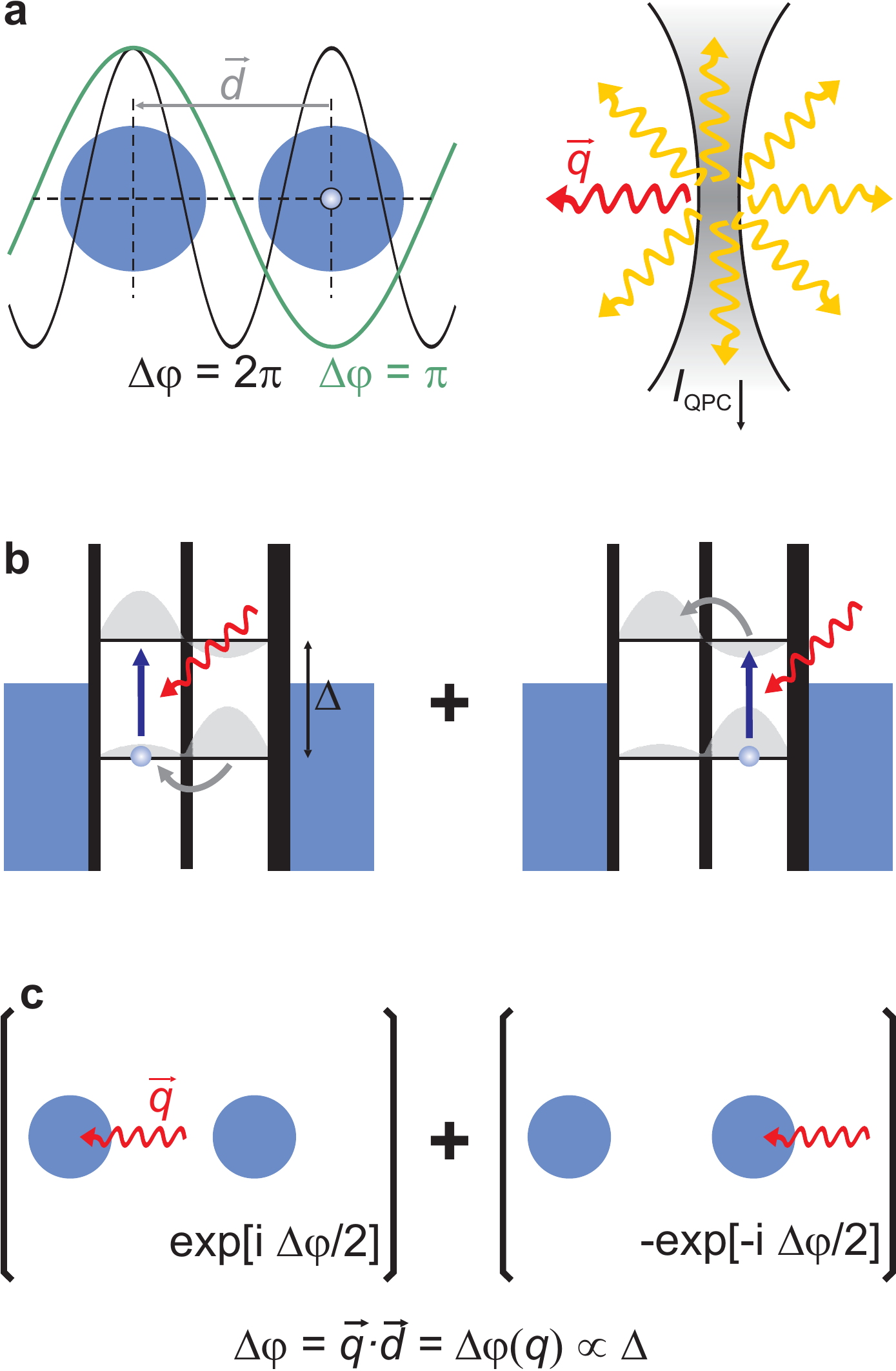}
\caption{\footnotesize
\textbf{Interference in quantum dot--phonon interactions. a,}
	 The back-action charge fluctuations of a QPC (right) used to measure quantum dots
	 generates locally non-equilibrium phonons (yellow and red squiggly lines); phonons emitted in the correct direction (red)
	 can travel from the QPC to the dots and excite them from their ground state.
	 In a semiclassical picture, the displacement wave associated with an
	 excited phonon mode will have a maximal effect
	 when it is exactly out-of-phase at the two dot sites (as shown in green), as it will cause an
	 oscillation in the effective energy detuning between the dots.
	 The relative phase of the wave between the dots is
	 $\Delta \varphi = \vec{q} \cdot \vec{d}$, where $\vec{q}$ is the phonon wavevector, and
	$\vec{d}$ is the vector connecting the two dot centers; constructive interference occurs when
	$\Delta \varphi = (2 n + 1) \pi$ with $n$ an integer.  In contrast, a minimal effect is expected
	when the displacement wave is in-phase at the two dot sites (as shown in black).
\textbf{b, }
	In a fully quantum description, absorption of a single phonon of wavevector $\vec{q}$ can occur
	via either the right or left dot; the amplitudes for each process add coherently to determine the final excitation probability.  As ground and excited states can have different electronic probability distributions (indicated in gray), excitation leads to a measurable change in the current through the QPC charge detector. The right barrier is very opaque (thick vertical line) to suppress tunneling between the right dot and the right lead. The blue arrow indicates the excitation while the grey arrow indicates the charge transfer between dots.
\textbf{c, }
	Schematic showing the two interfering processes for absorption of back-action generated
	phonons (red squiggly lines) by a DQD. The relative phase between the amplitude of each process is $\pi + \Delta \varphi$ (c.f. Eq.~(\ref{eq:HintDQD})). Note that in a given transition, the magnitude of $\vec{q}$ is determined by the energy splitting $\Delta$ between ground and excited states, while the direction of $\vec{q}$ is largely determined by the placement of the QPC with respect to the DQD axis. Note that an analogous interference effect involving {\it photon} absorption is not possible in our system, as the wavelength of a resonant photon would far exceed the size of the nanostructure.
}
\label{fig:Schematics}
\end{figure}

The back-action process considered in this paper involves deleterious inelastic tunneling events between two adjacent dots in a serial double or triple quantum dot (DQD, TQD).  The energy difference $\Delta$ between the initial and final electronic dot states is provided by the absorption of a non-equilibrium acoustic phonon, which itself is generated by the quantum point contact (QPC) detector \cite{Schinner2009}.  Such an absorption process between adjacent dots is constrained by the energy conservation condition  $\Delta = \hbar |\vec q| v_\mathrm{ph}$ ($v_\mathrm{ph}$ is the sound velocity, $\vec q$ the phonon wavevector).   More subtly, it is also sensitive to the difference in phase, $\Delta\varphi = \vec{d} \cdot \vec{q}$, of the associated phonon wave between the two dot positions, with $\vec{d}$ being the vector connecting the two dot centers \cite{Miller1960, Imry69}.  This $\vec{q}$ (and hence $\Delta$) dependent phase difference controls the matrix element for phonon-absorption since it determines whether the electron-phonon couplings in each of the two individual dots add constructively or destructively (see Fig.~\ref{fig:Schematics}) \cite{Brandes2005}. The result is an oscillatory probability for inelastic electron-transfer events involving phonon-absorption, with constructive interference occurring when $\Delta\phi=(2n+1)\pi$ (where $n$ is an integer).

\begin{figure}
\includegraphics[width=0.8\columnwidth]{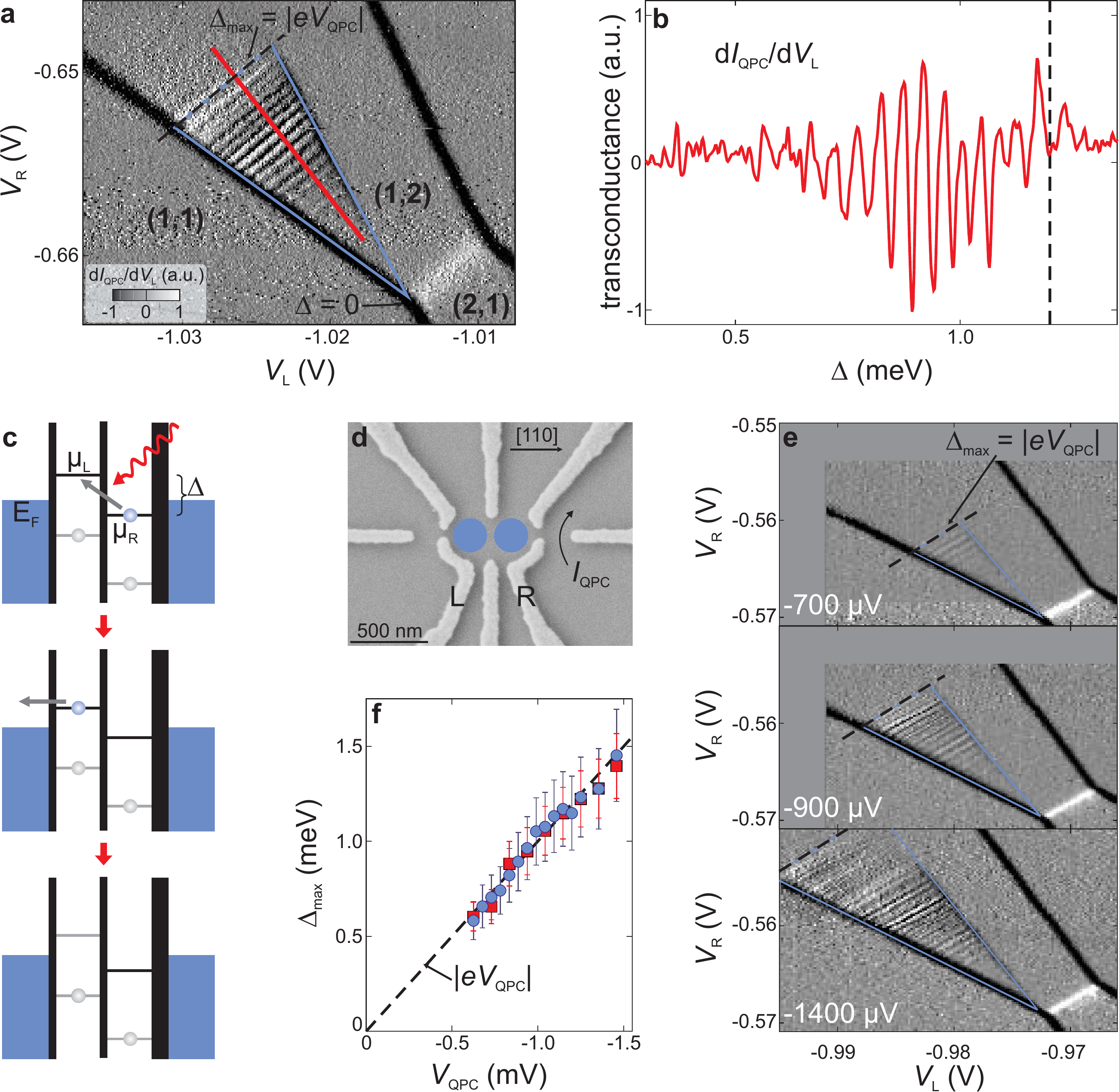}
\caption{\label{fig:DQD}\footnotesize \textbf{Interference in back-action in a DQD. a,} Charge stability diagram of a DQD at $\VQPC = -1200$ $\mu$V, showing the differential transconductance $\text{d}I_\text{QPC}/\text{d}V_\text{L}$ (in arbitrary units) numerically derived from the dc-current $I_{\text{QPC}}$ as a function of the voltages applied to gates ``L'' and ``R''. The number of electrons occupying the left and right dots is shown as ($N_\text{L},N_\text{R}$) for ground state configurations. At the charging lines (dark; local transconductance minima) the overall charge $N_\text{L}+N_\text{R}$ changes by one electron and at the charge transfer lines (white, local maxima) one electron moves between the dots. A triangle of back-action contains regular stripes which correspond to oscillations between the metastable configuration $(1,1)$ and the ground-state configuration $(1,2)$.
\textbf{b,} Trace along the red line in \textbf{a}.
\textbf{c,} Sketch of a possible back-action--induced excitation process $(1,2)\to(2,1)\to(1,1)$ for $(1,2)$ being the ground state configuration and the right tunnel barrier being closed (thick black vertical beam).
\textbf{d,} Scanning electron micrograph of the sample structure, the approximate positions of the dots are marked in blue. The crystalline direction [110] is marked by an arrow.
\textbf{e,} Three charge stability diagrams for $\VQPC = -700$ $\mu$V, $\VQPC = -900$ $\mu$V, and $\VQPC = -1400$ $\mu$V. The blue lines enclose the total striped area which increases proportionally to $\VQPC$. The dashed lines in \textbf{a}, \textbf{b}, and \textbf{e} correspond to $\Delta=\left|e\vqpc\right|$.
\textbf{f,} Detuning $\Delta_\mathrm{max}$ corresponding to the size of the striped triangle as a function of $\VQPC$ for two different series of measurements for differently tuned DQD systems (data in \textbf{e} belong to the blue circles). The error bars contain both the uncertainty of determining the voltage-to-energy conversion (see Methods section) as well as the error in triangle size from the measured data. The straight line denotes $\Delta_\mathrm{max}=\left|e\VQPC\right|$.}
\end{figure}

Data showing a pronounced back-action effect are shown in Fig.\ \ref{fig:DQD}a, which displays the stability diagram measured in charge detection for a few-electron DQD without a voltage drop between its left and right leads. The charge configuration of the quantum dot structures influences the conductance of a nearby QPC because of the capacitive coupling between the dots and the QPC. In order to serve as a charge detector it is necessary to drive a current through the detector QPC which, in turn, leads to the observed detector back-action. Multiple gates fabricated $\SI{85}{\nm}$ above a high-mobility two-dimensional electron system (2DES) are used to define two dots and two QPCs (Fig.\ \ref{fig:DQD}d). The differential transconductance $\dif I_\text{QPC}/\dif V_\mathrm{L}$ of the \textit{biased} charge detector QPC ($\VQPC=-1.2$\,mV) is plotted as a function of control gates $V_\text{L},V_\text{R}$. It shows local extrema at the boundaries between regions of different electronic ground states, yielding dark ``charging'' and white ``charge transfer'' lines. Specific ground state configurations are labeled ($N_\text{L},N_\text{R}$), where the integer $N_\alpha$ denotes the number of electrons in dot $\alpha = $ L (left) and R (right). As our DQD is cooled to $T\simeq30$\,mK the unmeasured DQD is expected to be in its ground state.

Detector back-action manifests itself within a distinct triangular-shaped region of deviations from the ground state configuration (1,2), where a pronounced pattern of repeated, parallel stripes is present. It indicates an oscillating probability to find the DQD in the excited configuration (1,1). The excitation process sketched in Fig.\ \ref{fig:DQD}c includes an inelastic tunneling transition $(1,2)\to(2,1)$ mediated by the absorption of a phonon, followed by an elastic (and therefore quick) tunneling process $(2,1)\to(1,1)$. In our measurements, the tunnel barrier between the right dot and right lead is tuned to be almost closed (see Figs.~\ref{fig:DQD}c,d). The direct transition $(1,1)\to(1,2)$ back into the ground state via an elastic tunneling process from the right lead is consequently very slow and the excited configuration (1,1) is metastable. The associated three-level dynamics can result in average non-thermal occupations \cite{Harbusch2010}. In this way a metastable excited state is essential to directly observe detector back-action in a low bandwidth stability diagram measurement. It requires asymmetric dot-lead tunnel couplings in case of a DQD (cf.\ Supplementary Information).

\begin{figure}
\includegraphics[width=0.85\columnwidth]{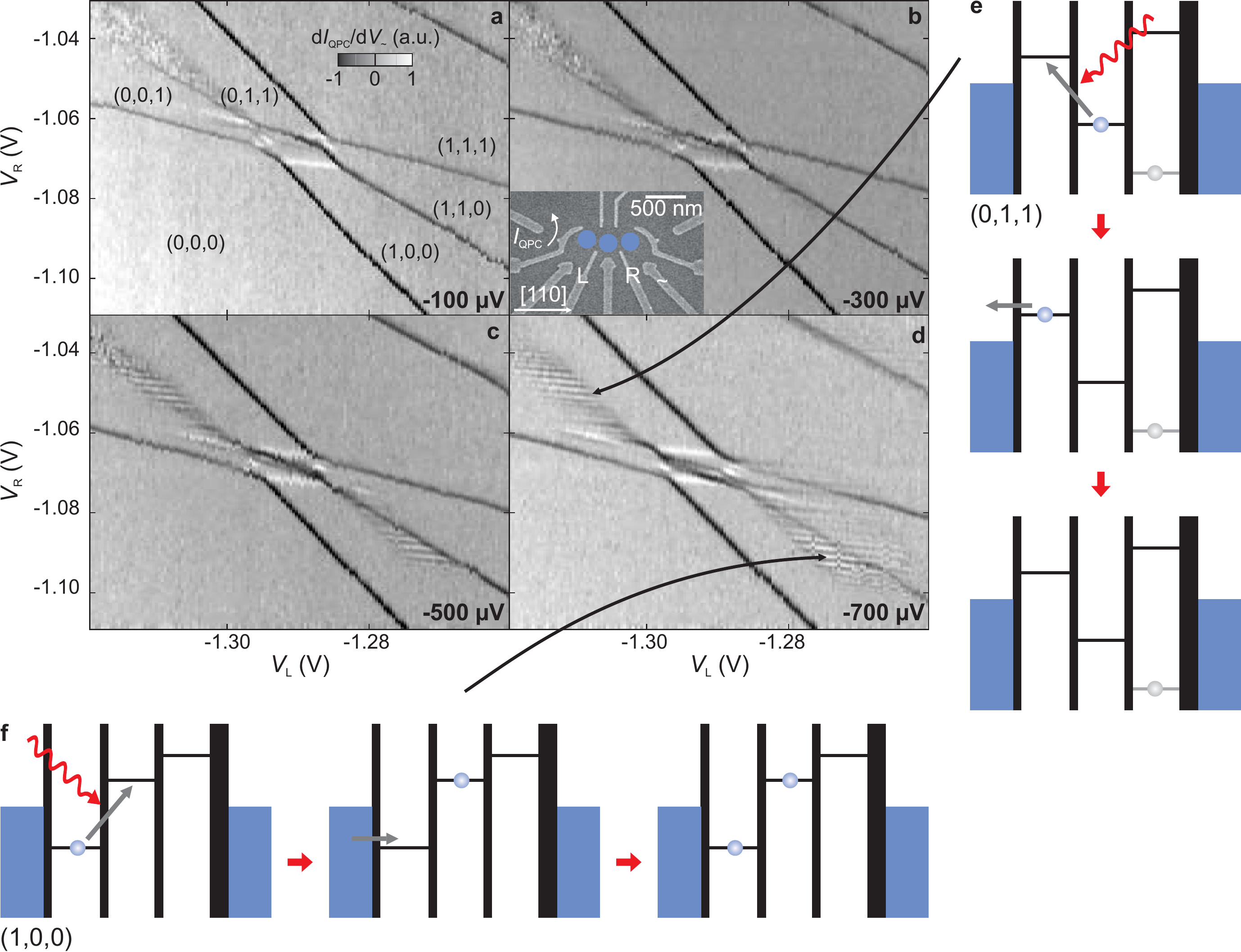}
\caption{\label{fig:TQD}\footnotesize\textbf{Interference in back-action in a TQD.}
\textbf{a--d,} Charge stability diagrams of a serial triple quantum dot (differential transconductance $\text{d}I_{\text{QPC}}/\text{d}V_\sim$ as in Fig.\ \ref{fig:DQD}). The linear response of $I_{\text{QPC}}$ is measured while the voltage applied to the gate marked with ``$\sim$'' is modulated by $\delta V_\sim=1\,$mV rms at the frequency of $f=13\,$Hz; inset of Fig.\ \ref{fig:TQD}b: scanning electron micrograph of the sample structure. Blue dots mark approximate positions of the quantum dots. The number of electrons occupying the left, central, and right dots is shown as ($N_\text{L},N_\text{C},N_\text{R}$) for ground state configurations. The QPC is biased by $V_\text{QPC}=-100$, $-300$, $-500$, and $-700\,\mu$V for \textbf{a}--\textbf{d}. Additional features are deviations from the ground state configuration caused by detector back-action.
\textbf{e,} Sketch of a possible back-action--induced process producing the striped pattern marked by the upper arrow in the (0,1,1) region in d. A phonon is absorbed and transfers an electron from the central to the left dot. Subsequently, the electron tunnels to the left lead, lowering the overall energy (of the system including dots and leads). The resulting configuration $(0,0,1)$ is a metastable excited state. \textbf{f,} Sketch of the back-action process in the $(1,0,0)$ ground state region (lower arrow in d), which is also based on the absorption of a phonon, a short-lived intermediate, and a metastable excited state.  The transitions in \textbf b and \textbf f are analogous to the process described in Fig.\ \ref{fig:DQD}c, with the closed tunnel barrier replaced by the rightmost dot in Coulomb blockade.}
\end{figure}

The stripe pattern constitutes the key signature of the coherent phonon-mediated back-action effect. It indicates that the probability to be in the excited configuration $(1,1)$ oscillates as a function of the energy detuning $\Delta$ between the intermediate state (2,1) and the ground-state configuration $(1,2)$  (see Fig.~\ref{fig:DQD}b); each stripe is thus parallel to the white charge transfer line where these states are degenerate (i.\,e.\ $\Delta = 0$; marked). The striped region is bounded by a line $\Delta = \Delta_{\rm max}$, indicating that there is a maximum energy available to excite the DQD. By seeing how this boundary changes with increasing $\VQPC$ (Figs.~\ref{fig:DQD}e--f), we find that $\Delta_{\rm max} \simeq \left|e\VQPC\right|$, consistent with the QPC indeed being the energy source for the initial DQD excitation.

The geometry of the back-action regions as well as the influences of temperature and the orbital excitation spectrum are discussed in the Supplementary Information. In short, the remaining boundaries of the triangle-shaped regions of back-action correspond to energy thresholds for lead tunneling.
The width of each stripe is largely independent of temperature; this is indicative of an excitation process involving electron transfers between dots, without any involvement of lead electrons (see Fig.~\ref{fig:DQD}c and Figs.\ \ref{fig:TQD}e,f).
The regular spacing of the stripe features in both DQD and TQD (discussed below) experiments over so many stripes eliminates the possibility that they are due to resonances with orbital excitations of a dot, as there is no reason to expect such a uniform level spacing; further, the energy spacing between the stripes is much smaller than would be expected for the average level spacing of the small dots studied here.

\begin{figure}
\includegraphics[width=0.85\columnwidth]{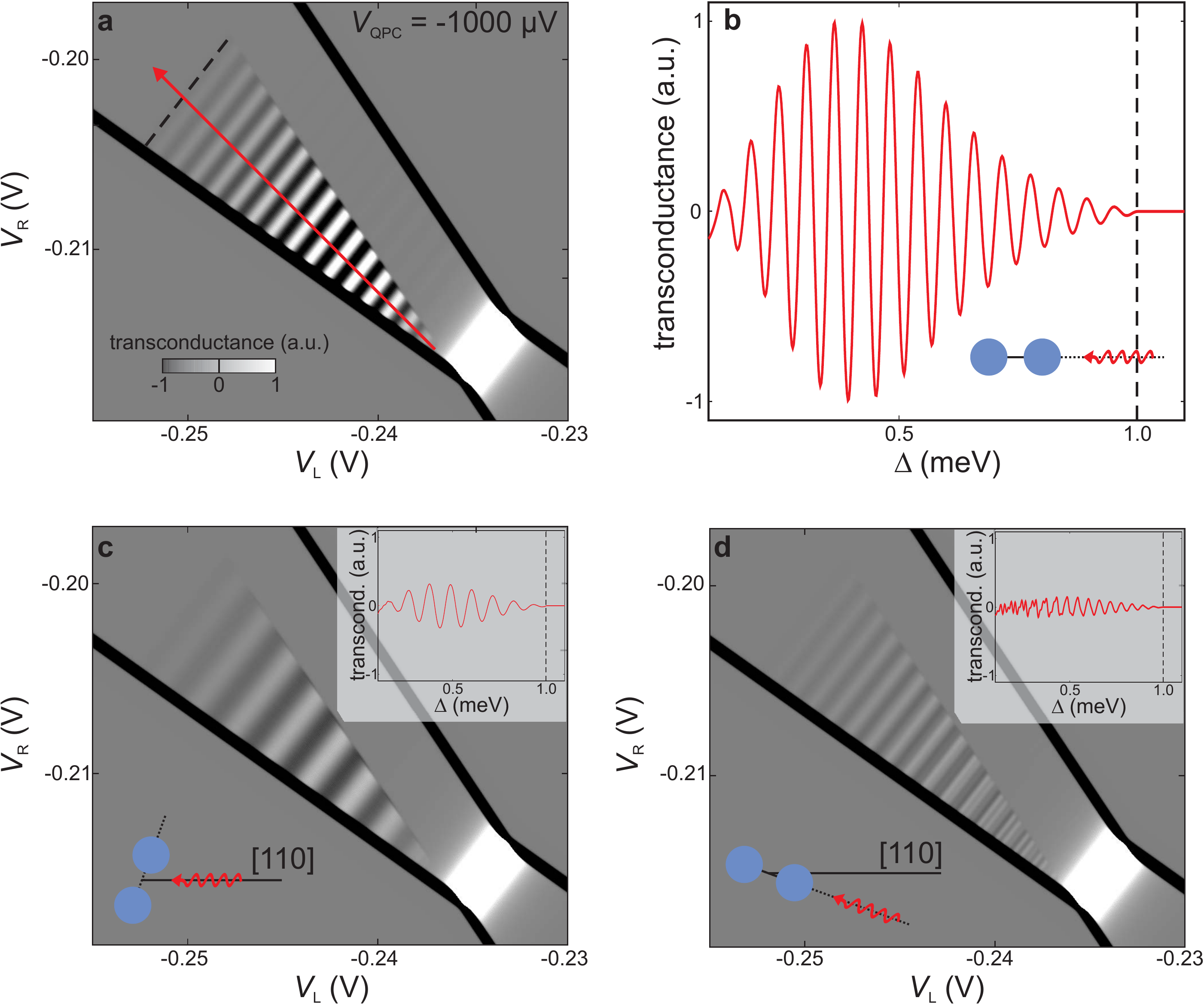}
\caption{\footnotesize
\textbf{Theoretical results.}
\textbf{a, }
	 DQD stability diagram calculated using the theory described in the main text, which describes the emission of acoustic phonons by the non-equilibrium QPC charge fluctuations, and their coherent absorption by the double dot. Plotted is the derivative of the weighted average DQD charge $\langle n_R \rangle + \epsilon \langle n_L \rangle$ ($\epsilon \simeq 0.4$, see Methods section) with respect  to the gate voltage $V_L$ in the DQD experiment, as a function of the gate voltages $V_L$ and $V_R$ (see Fig. \ref{fig:DQD}); this is proportional to the measured transconductance of the QPC.  Parameters are taken to be the same as those in the DQD experiment; in particular, the dots and QPC are taken to be collinear and aligned with the $[110]$ crystallographic axis. For this orientation, one finds that the so-called ``fast transverse'' acoustic phonon mode \cite{Jasiukiewicz:1998} makes the dominant contribution. Pronounced stripe patterns are seen, similar to the experiment.
\textbf{b, }
	A cut of the calculated stability diagram in a), along the indicated line.  The suppression of oscillations at low values of ground state--excited state energy detuning $\Delta$ is the result of screening, and is consistent with experiment. The oscillations are cut off at $\Delta = \Delta_{\rm max} \sim \left|eV_{\rm qpc}\right|$, also in agreement with experiment.
\textbf{c, }
	Calculated stability diagram for identical parameters as a), except that the QPC is not collinear with the two dots (the QPC-dot axis is rotated $70^{\circ}$ from the dot-dot axis $\vec{d}$). The result is a suppression of the back-action-induced stripe pattern.
\textbf{d, }
	Calculated stability diagram for identical parameters as a), except that the orientation of the DQD-QPC axis is now rotated $20^\circ$ from the $[110]$ direction. The resulting suppression of interference is due to the anisotropy of electron-phonon interactions:  the effective electron-phonon interactions are weaker in this direction. The change in orientation also causes the ``slow transverse'' acoustic phonon mode to contribute, yielding a second, high-frequency oscillation. Insets in \textbf{c} and \textbf{d} show cuts through the stability diagram of the transconductance along the same line indicated in red in \textbf{a}.
}
\label{fig:Theory}
\end{figure}

To quantify the interpretation of the stripe pattern in Fig.~\ref{fig:DQD} in terms of interference and QPC back-action, we have developed a theoretical model which describes the generation of phonons by the non-equilibrium QPC charge fluctuations \cite{Young2010}, and their coherent absorption by the DQD. These fluctuations represent the fundamental back-action of the measurement-- their magnitude is bounded from below by the rate at which information is obtained from the QPC via a Heisenberg-like inequality \cite{Young2010}.  Given this, the back-action charge noise mechanism we describe must necessarily make a contribution to the observed oscillations.  This mechanism is also consistent with the high visibility of the oscillations, as such visibility requires a highly localized source of hot phonons.  While we cannot completely rule out that other, less direct back-action mechanisms contribute additionally (e.g. generation of hot phonons in the QPC leads), it is not clear that such mechanisms would also yield such high-visibility oscillations. We describe bulk acoustic phonon modes of GaAs interacting with both electrons in the DQD, as well as with the fluctuating charge density of the biased QPC via a screened piezoelectric interaction.  Using Keldysh perturbation theory, we can calculate the DQD state in the presence of back-action (see Fig.~\ref{fig:Theory} and Supplementary Information). The relevant part of the dots-phonon interaction (i.e.~terms that can cause transitions in the dot) take the form:
\begin{equation}
	\hat{H}_{\rm int} =   \frac{t_c}{\Delta} \sum_{\vec{q}, \mu}
	\lambda_{\vec{q},\mu} \left(  e^{i \vec{q} \cdot \vec{r}_L} -  e^{i \vec{q} \cdot \vec{r}_R}   \right)
	\left(   \hat{a}^{\phantom{\dagger}}_{\mu, \vec{q}} + \hat{a}^\dagger_{\mu,-\vec{q}}   \right)
	\Big( |g\rangle \langle e| + h.c. \Big)
	\label{eq:HintDQD}
\end{equation}
Here $|g\rangle$ ($|e \rangle$) denotes the DQD ground (excited) state, $t_c$ is the pertinent interdot tunnel matrix element, $\hat{a}_{\mu, \vec{q}}$ destroys a phonon of wavector $\vec{q}$ in branch $\mu$, and $\lambda_{\vec{q},\mu }$ is the effective matrix element (screened) for the interaction of phonons with a single dot.  The first bracketed factor in the sum of Eq.~(\ref{eq:HintDQD}) denotes the key interference of relevance: the two terms correspond to phonons interacting with electrons in either the left or right dot
(which are centered at $\vec{r}_L$ and $\vec{r}_R$ respectively), see Fig.~\ref{fig:Schematics}b.

Despite the explicit interference evident in Eq.~(\ref{eq:HintDQD}), geometric averaging can still strongly suppress interference oscillations in observable quantities. Simply put, while the DQD ground-excited energy splitting fixes the {\it magnitude} of a phonon participating in an inelastic tunneling event, it does not specify its direction; hence, the relative phase in the first term of Eq.~(\ref{eq:HintDQD}) is not completely determined by $\Delta$. This is typically the case in situations probing the {\it emission} of acoustic phonons by {\it biased} DQDs \cite{Brandes2005}, where interference oscillations are observed, albeit with much smaller visibilities than seen here \cite{Fujisawa1998,Roulleau2011}. In contrast, the simple geometric filtering depicted in Fig.~\ref{fig:Schematics}a suggests that this averaging need not play a role in phonon absorption, as only phonons traveling from the QPC to the dots contribute.  This is supported by our theoretical calculations, which also exhibit strong oscillations for realistic parameter values, and show a pronounced enhancement of interference oscillations when the DQD and QPC are all collinear (see Fig.~\ref{fig:Theory}a-c).

The theory is also able to capture other aspects of the experimental data: in particular, the size of the back-action triangle grows with $\left| \vqpc \right|$, and the lowest-energy stripes (i.\,e.\ smallest values of $\Delta$ corresponding to long phonon wavelength) are suppressed due to screening
effects  (see Fig.~4b).  Using the fact that in the experiment the QPC and DQD are approximately collinear, the measured spacing of the interference parameter $\delta \Delta = 45 \mu$eV in the DQD data of Fig.~\ref{fig:DQD}d yields a DQD separation $d = h v_s / \delta \Delta \simeq 250$ nm; this is in good agreement with the separation estimated from SEM images (Fig.~\ref{fig:DQD}d).  The theory also shows that due to the anisotropy of the electron-phonon matrix elements $\lambda_{\vec{q},\mu}$, the overall magnitude of the phonon-induced back-action is sensitive to the orientation of the dots-QPC axis with respect to crystallographic axes.  This dependence on orientation is demonstrated in Fig.~\ref{fig:Theory}d.  More details on the theoretical treatment is provided in the Methods section and
Supplementary Information.

While we have focused so far on back-action in DQDs, the mechanism we describe is extremely general, and is in fact even more ubiquitous in systems with more than two dots.  As discussed, a key requirement to see the effect is the existence of a long-lived metastable excited state.  Such a situation occurs rather naturally in serial TQD structures \cite{Gaudreau2006,Schroer2007,Rogge2008,Rogge2009,Granger2010}, as the center dot is effectively decoupled from one of the leads whenever either one of the other two dots (left, right) is in Coulomb blockade.  This directly yields a metastable excited state in which the charge of the middle dot is unable to relax. As a consequence deviations from the ground state configuration are often observed along the charging line of the center dot \cite{Schroer2007} and back-action effects occur naturally in the stability diagram. We study detector back-action in a TQD in Fig.\ \ref{fig:TQD} by successively increasing $\left|e\vqpc\right|$. Already at relatively small bias $|\VQPC|\le300\, \mu$V (Figs.\ \ref{fig:TQD}a and b) a triangular-shaped region of telegraph noise is observed along the central charging line \cite{Schroer2007}. It indicates slowly fluctuating deviations from the ground state configuration, which can be caused by external noise or detector back-action \cite{Taubert2008}. The underlying excitation processes, sketched in Figs.~\ref{fig:TQD}e--f, are similar as the one discussed above for the DQD. Indeed, the population of the right dot does not fluctuate; it takes the same role as the closed barrier in case of the DQD, namely to block charge exchange between the center dot and the right lead. Further increasing $|\VQPC|$ to $500\,\mu$V in (Fig.\ \ref{fig:TQD}c) reveals the familiar pattern of equally spaced stripes both within the $(1,0,0)$ and $(0,1,1)$ regions. As the bias is increased even more to $\left|\VQPC\right|=700\,\mu$V (Fig.\ \ref{fig:TQD}d) the striped regions expand further, revealing the $\VQPC$ dependence also observed in case of the DQD (see Fig.\ \ref{fig:DQD}f).

By considering experimental data on both DQD and TQD systems, we have demonstrated that interference can strongly affect the phonon-mediated back-action generated by a QPC in quantum dot circuits.  Further, we have shown that this effect is well described by a basic theoretical model incorporating the generation of phonons by the QPC detector and their coherent absorption by the dots.  Our study suggests the possibility of mitigating back-action effects by making use of this interference.
 One could, for example, endeavour to first tune the DQD/TQD to an operating point where destructive interference suppresses phonon absorption, and only then energize the QPC to make a measurement.
 More complex schemes which also incorporate the anisotropy of the electron-phonon interaction with respect to crystallographic axes could potentially yield even greater back-action reduction. Since the piezoelectric coupling to in-plane phonons is maximized in the $\langle 110 \rangle$ directions \cite{Jasiukiewicz:1998}, by aligning the QPC-DQD axis away from these directions, one could appreciably decrease the phonon-mediated back-action excitation discussed here (see e.g. Fig. 4a vs 4d).

\section{Methods}

{\bf Experiment. }The samples have been fabricated from GaAs\,/\,AlGaAs heterostructures containing 2DESs 100\,nm (TQD) and 85\,nm (DQD) below the surface, respectively.
The 2DESs are characterized at cryogenic temperatures by the carrier densities of $n_S = 2.1 \times 10^{11}  \, \rm{cm}^{-2}$ and
$n_S = 1.9 \times 10^{11}  \, \rm{cm}^{-2}$ with the mobilities of $\mu = 1.72 \times 10^6 \, \rm{cm}^2 / \rm{V s}$ and $\mu = 1.19 \times 10^6 \, \rm{cm}^2 / \rm{V s}$ for the TQD and DQD, respectively.
Metallic gate electrodes have been fabricated on the sample surface by electron-beam lithography and standard evaporation/liftoff techniques. Negative voltages applied to these electrodes are used to locally deplete the 2DESs and thereby define the quantum dot and QPC structures. All measurements have been performed in dilution refrigerators at cryogenic temperatures below $100\,$mK. To detect the charge configuration of the TQD the voltage of one gate of the TQD was slightly modulated and the detector differential transconductance was measured in linear response (ac set-up). In addition a constant voltage was applied across the QPC to enhance detector back-action. In case of the DQD only a constant voltage was applied across the QPC and the direct current $I_\text{QPC}$ flowing through the QPC was measured to detect the charge configuration of the DQD (dc set-up). The differential transconductance $\dif I_\text{QPC}/\dif V_\text{L}$ was then computed numerically. Both methods result in the differential transconductance and their interpretation is identical.

To interpret the observed back-action in terms of the energy detuning $\Delta$ between different charge configurations an accurate conversion from gate voltages to units of energy is necessary. Such a linear transformation has been performed following the methods described in \cite{Taubert2011}. The conversion relation reads $\Delta = \left( \argl - \algl \right) \vl + \left( \argr - \algr\right) \vr$, with the following set of conversion factors determined for the red symbols in Fig.\ \ref{fig:DQD}f:
\argl = (54 $\pm$ 5) \SI{\milli\electronvolt\per\volt}, \argr = (105 $\pm$ 4) \SI{\milli\electronvolt\per\volt}, \algr = (62 $\pm$ 4) \SI{\milli\electronvolt\per\volt}, \algl = (90 $\pm$ 5) \SI{\milli\electronvolt\per\volt}. The conversion factors related to the blue symbols in Fig.\ \ref{fig:DQD}f read \argl = (65 $\pm$ 6) \SI{\milli\electronvolt\per\volt}, \argr = (109 $\pm$ 7) \SI{\milli\electronvolt\per\volt}, \algr = (61 $\pm$ 6) \SI{\milli\electronvolt\per\volt}, \algl = (97 $\pm$ 7)  \SI{\milli\electronvolt\per\volt}.

{\bf Theory.  }The fluctuating non-equilibrium QPC charge density operator $\hat{\rho}(\vec{r})$ is modelled as
$\hat{\rho}(\vec{r}) = f(\vec{r}) \hat{Q}$, where the total charge operator $\hat{Q}$ is described by scattering theory (cf. Refs.~\cite{Pedersen1998,Young2010}). Note that as we are interested in a single-channel QPC, the spatial profile $f(\vec{r})$ of the fluctuating QPC charge density is fixed; for simplicity, we take it to be a Gaussian of width  $r_{\rm QPC}$. This fluctuating QPC charge density is coupled to acoustic phonons via the standard piezoelectric interaction (using parameters appropriate for GaAs \cite{Jasiukiewicz:1998}).  We calculate the Keldysh Green functions of the acoustic phonons in the presence of this coupling to the QPC, working to leading order in the electron-phonon interaction, and using scattering theory to calculate the QPC Keldysh Green functions.  We then use these ``dressed" phonon Green functions to calculate the Fermi Golden rule excitation rate of the DQD via the coupling described in Eq.~(\ref{eq:HintDQD}). This excitation rate is finally incorporated into a master equation describing the occupation probability of the three relevant DQD states (see Fig.~\ref{fig:DQD}c). In addition to the excitation rate (top panel of Fig.~\ref{fig:DQD}c), there is a rate $\Gamma_{\rm fast}$ describing the tunneling from the excited state to the metastable auxiliary state (middle panel of Fig.~\ref{fig:DQD}c), and a rate $\Gamma_{\rm slow}$ describing the slow decay back to the true ground state (bottom panel of Fig.~\ref{fig:DQD}c). We take $\Gamma_{\rm fast} = 1\,$GHz and  $\Gamma_{\rm slow} = 10\,$kHz; in this regime of $\Gamma_{\rm fast} \gg \Gamma_{\rm slow}$, the non-ground state population of the DQD is independent of $\Gamma_{\rm fast}$, whereas  $\Gamma_{\rm slow}$ determines the overall magnitude of the interference oscillations. By using the master equation to calculate the stability diagram as a function of gate voltages, one can obtain the DQD charge susceptibility $d ( \langle n_R \rangle + \epsilon \langle n_L \rangle) / d V_L$ , which is proportional to the measured transconductance.  The parameter $\epsilon \sim 0.4$ characterizes the QPC's asymmetric response to charge in the R versus L dot.  Further details about the explicit form of $\lambda_{\vec{q},\mu}$ (including the role of screening and dimensionality)  are provided in the Supplementary Information.

\section{Author contributions}
D.\ T.\ fabricated the DQD samples, performed the DQD experiments and analysed the data. D.\ H.\ performed preliminary experiments on another DQD sample. A.\ K.\ fabricated the TQD sample. L.\ G., G.\ G., and S.\ S.\ performed the TQD experiments and analysed the data. P.\ Z.\ assisted in these experiments. D.\ S.\ and W.\ W.\ grew the heterostructures for the DQD samples; Z.~R.~W.\ optimized and grew the heterostructure for the TQD sample. D.\ T., L.\ G., C.\ E.\ Y., A.\ A.\ C., A.\ S.\ S., and S.\ L.\ wrote the paper. C.\ E.\ Y.\ and A.\ A.\ C.\ developed the theoretical model and supported both experimental groups. A.\ S.\ S.\ and S.\ L.\ supervised the experimental collaboration from Ottawa and Munich.

\section{Acknowledgments}

S.\ L.\ and D.\ T.\ acknowledge financial support by the German Science Foundation via SFB 631, LU 819/4-1, and the German Excellence Initiative via the ``Nanosystems Initiative Munich'' (NIM). G.\ G.\ acknowledges funding from the NRC-CNRS collaboration.  A.\ S.\ S.\ and A.\ C.\ C.\ acknowledge funding from NSERC and CIFAR.


\newpage
\begin{center}
\section{\underline{SUPPLEMENTARY INFORMATION}}
\end{center}

\section{The origin of detector back-action--induced nonequilibrium occupations}
\label{sub_back_origin}

The stability diagrams in Figs.\ 2 and 3 in the main paper are modified by the absorption of phonons emitted from the quantum point contact (QPC) which is used as a charge detector. In this section we detail the microscopic tunneling processes involved in examples of detector back-action and discuss the general conditions under which the observed situations of nonequilibrium charge configuration occupation occur.

The microscopic back-action processes involved in the triple quantum dot (TQD) are sketched in Figs.\ 3e,f. In the lower right region of back-action the ground state configuration is $(1,0,0)$. As shown in Fig.\ 3f the electron in the left dot is transferred to the center dot $(1,0,0)\rightarrow(0,1,0)$ via the resonant absorption of a phonon. Then, an additional electron tunnels from the left lead into the left dot $(0,1,0)\rightarrow(1,1,0)$. This latter process is a very fast resonant tunneling event since the lead provides a continuum of occupied states below the Fermi energy. However, the configuration $(1,1,0)$ is metastable since the relaxation transition $(1,1,0)\rightarrow(1,0,0)$ is hindered by Coulomb blockade.

In the upper left region of back-action the ground state configuration is (0,1,1). As shown in Fig.\ 3e, the electron in the center dot is transferred to the left dot $(0,1,1)\rightarrow(1,0,1)$ via the resonant absorption of a phonon. Following this the same electron tunnels from the left dot into the left lead $(1,0,1)\rightarrow(0,0,1)$. This is again a very fast resonant tunneling process since the lead provides a continuum of unoccupied states above the Fermi energy. However, the configuration (0,0,1) is metastable since the transition $(0,0,1)\rightarrow(0,1,1)$ is hindered by Coulomb blockade.

Note that in both situations described above, the charge of the right dot remains fixed. Charge exchange between the center dot and the right lead is always hindered by Coulomb blockade. In case of a double quantum dot (DQD) we consequently just replace the right dot by a large tunnel barrier, in order to suppress charge exchange between the right dot and leads. Indeed, the microscopic back-action process observed in Fig.\ 2 is identical (except for uninvolved electrons) to the second one described above and can be summarized by the two-stage transition $(1,2)\rightarrow(2,1)\rightarrow(1,1)$.

In our measurements we observe a strong nonequilibrium occupation and can even reach full population inversion, where the ground state remains unoccupied. The principles of thermodynamics permit such a situation, but only under two specific conditions: First, a nonequilibrium energy source is required (in thermal equilibrium the ground state has the largest occupation probability and the occupation of all states is predetermined by the Boltzmann factor). In our case the QPC charge detector acts as the nonequilibrium energy source. Secondly, transitions between at least three states must be involved (in nonequilibrium, population inversion is impossible in a two-level system because the emission rate is always larger than the absorption rate due to spontaneous emission). This second condition emphasizes the importance of the (short lived) intermediate state which decays rapidly into the metastable excited state. It is the Coulomb blockade effect which makes this excited state metastable and enables one to observe the detector back-action. If not all of these requirements are fulfilled the back-action effect might still exist but would not be directly observed within the stability diagram of a quantum dot circuit.

\section{Geometry of back-action regions in stability diagrams}
\label{sub_back_borders}

\begin{figure}[ht]
\includegraphics[width=0.9\columnwidth]{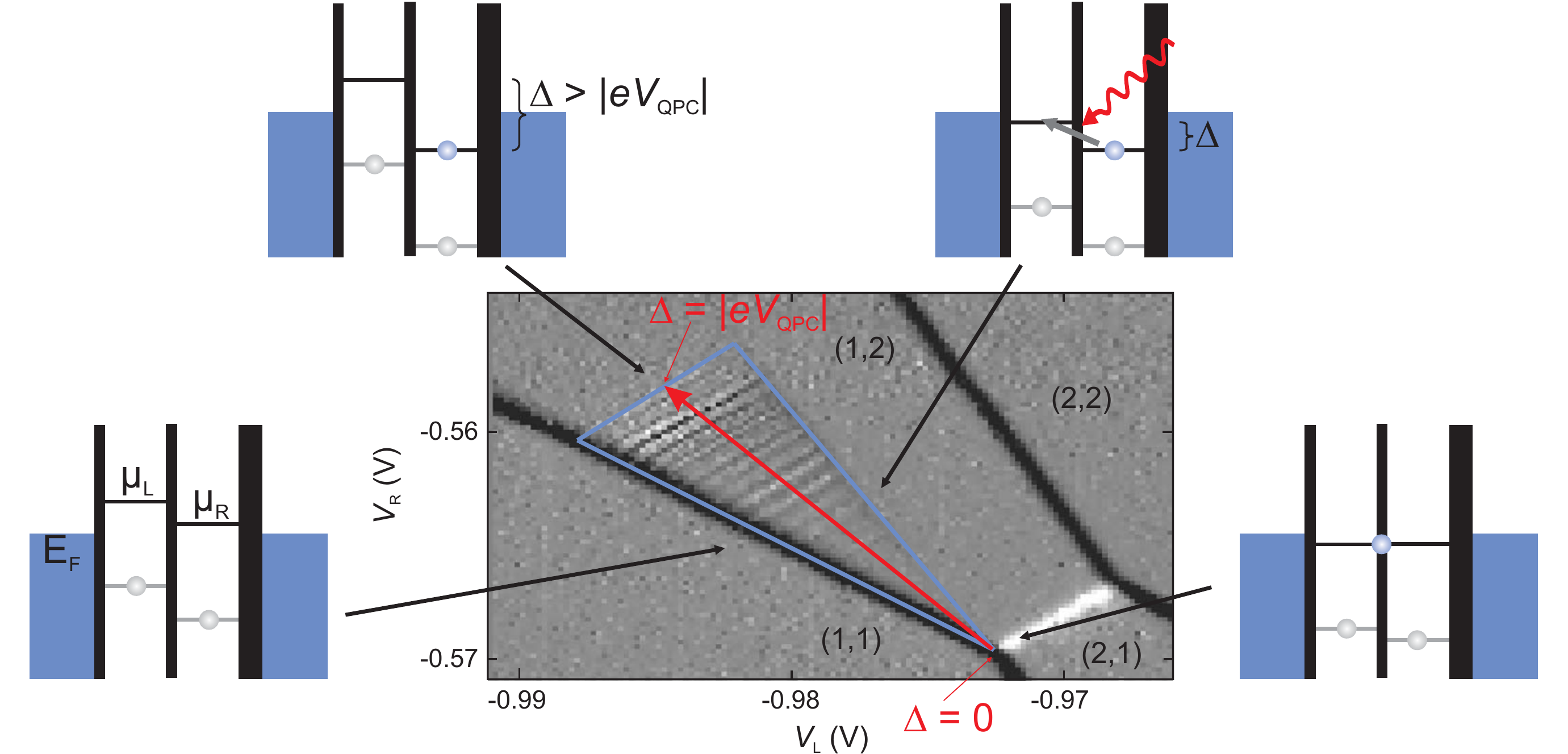}
\caption{\label{sup_borders}
\textbf{Geometry of back-action regions.}
Charge stability diagram of the serial DQD (excerpt from Fig.\ 2e) measured at $T_\mathrm{2DES}\simeq35\,$mK using charge detection with a QPC biased with $V_\mathrm{QPC}=\SI{-0.9}{\mV}$ (compare Fig.\ 2d). Each dot is charged by one or two electrons. The triangular-shaped region containing back-action--induced average deviations from the ground state occupation is framed by a blue line. The energy detuning $\Delta$ between configurations $(1,2)$ and $(2,1)$ along the red arrow spans the regime $0\le\Delta\le|eV_\mathrm{QPC}|$. The four sketches, illustrating the ground state configuration at selected points just outside of this triangle, are used to explain the geometry of the back-action region (see text). The phonon-mediated transition relevant for the observed interference pattern is also indicated in one sketch (cf.\ main paper). Blue areas depict occupied states in the degenerate 2DES of the leads with Fermi energy $E_\mathrm F$ while the vertical lines represent the tunnel barriers. Black horizontal lines mark the discrete chemical potentials $\mu_\mathrm L$ and $\mu_\mathrm R$ of the dots, the energy needed for adding the third electron, based on configuration $(1,1)$. The tunnel barrier between the right dot and the right lead (fat vertical line) is almost closed.}
\end{figure}

In the following we discuss the geometrical shape of the regions which contain back-action and interference patterns (parallel stripes). Fig.\ 2 of the main paper shows one back-action region for a DQD and Fig.\ 3 contains two distinct regions of back-action in the stability diagram of a TQD. As an example we replot in Fig.\ \ref{sup_borders} the charge stability diagram of the DQD for $V_\mathrm{QPC}=\SI{-0.9}{\mV}$ already shown in Fig.\ 2e. 
It shows the differential transconductance $\mathrm d I_\mathrm{QPC}/\mathrm d V_\mathrm{L}$ measured with a biased QPC which is capacitively coupled to the DQD (compare Fig.\ 2d); the black lines are charging lines while the white line is a charge reconfiguration line. As this measurement has been performed at very low temperature ($T_\mathrm{2DES}\sim35\,$mK), in thermal equilibrium the charge configuration of the DQD is well defined in the areas between these lines and corresponds to the ground state. The deviations from the ground state configuration (1,2) in a triangular-shaped region in Fig.\ \ref{sup_borders} are the result of detector back-action. The microscopic processes involved in this nonequilibrium phenomenon and the resulting interference pattern (parallel stripes) are discussed in detail in Sec.\ \ref{sub_back_origin} as well as the main paper.

Here, we specifically address the position and boundaries of the triangle. For this purpose it is sufficient to consider the transition $(1,2)\rightarrow(2,1)$ for which an energy detuning, $\Delta$, needs to be overcome. The tip of the triangle [on the triple point $(1,1)\leftrightarrow(2,1)\leftrightarrow(1,2)$] is characterized by $\Delta=0$ as in the lower right sketch in Fig.\ \ref{sup_borders}. The triangles base line, which lies parallel to the charge reconfiguration line $(1,2)\leftrightarrow(2,1)$, is determined by $\Delta\simeq|eV_\mathrm{QPC}|$ (compare upper left sketch and red arrow in Fig.\ \ref{sup_borders}). This range in $\Delta$ spans the full spectrum of energy quanta that the QPC can emit. On its left the triangle is bounded by the charging line of the right dot, $(1,1)\leftrightarrow(1,2)$. Below this line, (1,1) is the ground state configuration and the probability for (1,2) to be occupied is low (see lower left sketch in Fig.\ \ref{sup_borders}). As a consequence the transition $(1,2)\rightarrow(2,1)$ is suppressed and no back-action is observed. The right boundary of the triangle is a continuation of the charging line of the left dot $(1,1)\leftrightarrow(2,1)$. Above this line the configuration (2,1) has a lower energy than (1,1) (compare upper right sketch in Fig.\ \ref{sup_borders}). Here, the transition $(1,1)\rightarrow(2,1)$ occurs rapidly via resonant tunneling of an electron from the left lead into the left dot; the configuration (1,1) is no longer metastable, so the ground state configuration (1,2) is occupied most of the time and no back-action is observed.

The boundaries of the detector back-action--induced triangles in the TQD stability diagram in Fig.\ 3 can be explained similarly.

\subsection{Influence of the electronic excitation spectrum}

The back-action--induced nonequilibrium occupation within the triangles oscillates as a function of energy detuning $\Delta$. In the differential transconductance signal this leads to a regular pattern of stripes parallel to the relevant charge reconfiguration line (see, for instance, Figs.\ 2, 3 and 4 in the main paper). We interpret these occupation oscillations as an interference pattern of two competing phonon-absorption processes which enable an inelastic interdot electron tunneling transition. The period $\delta\Delta$ of the interference pattern corresponds to the energy of a phonon with a wavelength matching the distance between the two dots (see main paper). The regular spacing $\delta\Delta\sim50\,\mu$eV of the interference stripes excludes any interpretation in terms of the electronic excitation spectra of the dots, which are less regular with energy spacings much larger than $50\,\mu$eV. Note that small variations of the period $\delta\Delta$ are the result of a beating of different contributing phonon modes, details of which will be the focus of a future paper. Let us now discuss the role of the electronic excitation spectrum of the dots.

\begin{figure}[ht]
\includegraphics[width=1.0\columnwidth]{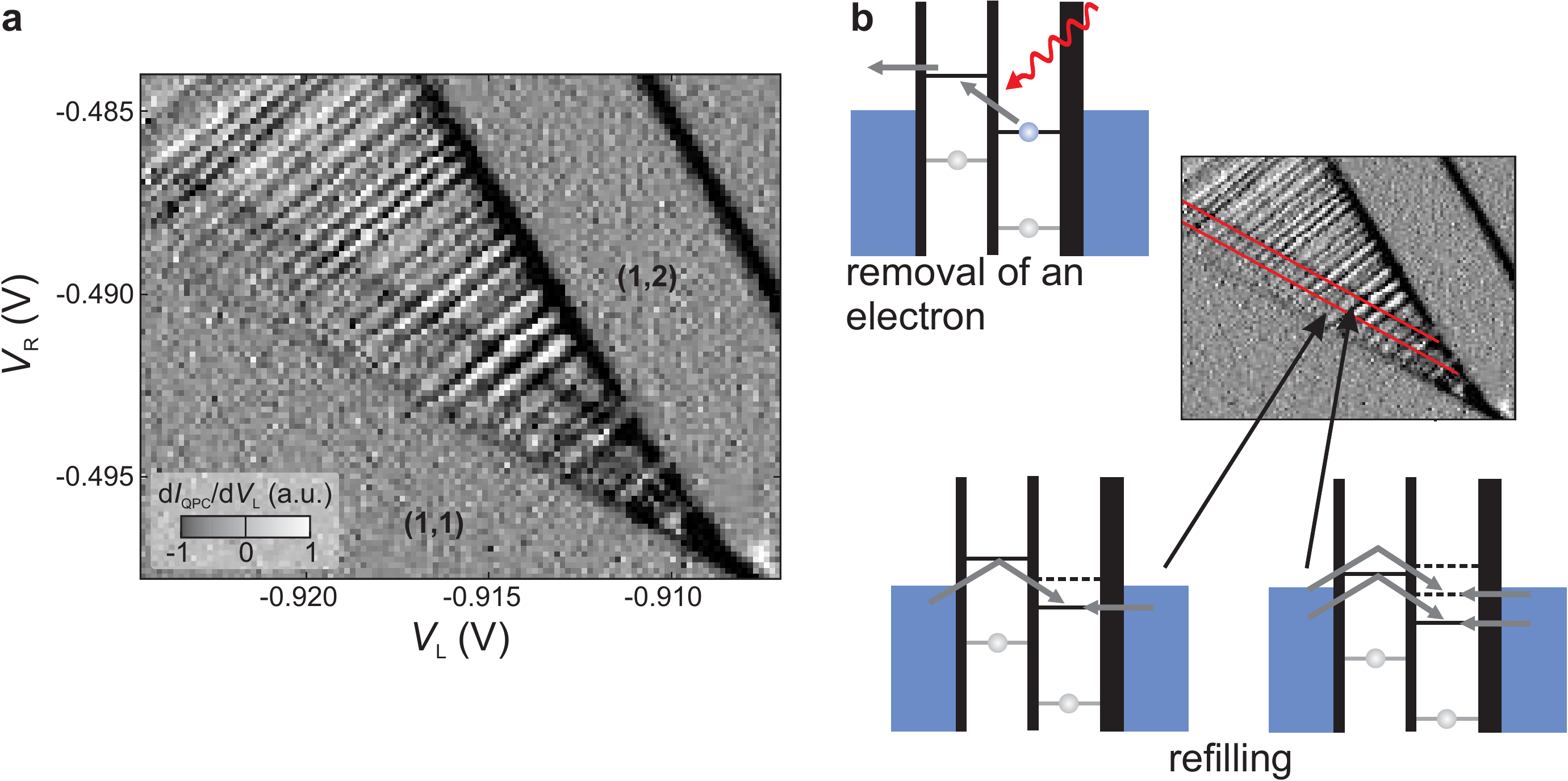}
\caption{\label{sup_excited}\textbf{Influence of the electronic excitation spectrum of the DQD.} \textbf{a,} Charge stability diagram measured with the DQD sample at $V_\mathrm{QPC}=-1.0\,$meV, similar to Fig.\ 2a. Two electronic excited states in the right dot cause additional dark lines of negative differential transconductance parallel to the left border of the striped triangle. \textbf{b,} Same plot as in \textbf{a}, with the additional lines marked in red. The upper sketch (as in Fig.\ \ref{sup_borders}) shows the phonon-mediated transition $(1,2)\rightarrow(2,1)$, which primarily involves the electronic ground states of the two dots and which is immediately followed by the quick transition $(2,1)\rightarrow(1,1)$ (compare Sec.\ \ref{sub_back_origin}). The other two sketches depict possible processes of the refilling transition $(1,1)\rightarrow(2,1)$ if only the ground state (one channel) or in addition an excited state contributes (two channels).
}
\end{figure}

Electronic excitations in the individual dots can influence the occupation probability of nonequilibrium configurations; an example is given in Ref.\ \cite{Harbusch2010}. In our experiments discussed here, the main effect of excited dot states is their influence on the refilling rate, which returns the system from the nonequilibrium configuration back to the ground state configuration. This influence can be directly seen in Fig.\ \ref{sup_excited}a
which plots a typical charge stability diagram in differential transconductance for the case when the barrier between the right dot and the lead is almost closed (compare sketches in Fig.\ \ref{sup_excited}b). Two dark lines, parallel to the charging line of the right dot, are visible in the back-action--induced triangle in Fig.\ \ref{sup_excited}a. They mark resonances of excited states of the right dot with the Fermi energy. The characteristic level spacing of the electronic spectrum is roughly \SI{120}{\ueV}, considerably larger than the phonon-induced oscillation period of $\delta\Delta\sim50\,\mu$eV. If the chemical potential of such an excited state falls below the Fermi energy (above the relevant resonance in Fig.\ \ref{sup_excited}a) an additional channel for the transition $(1,1)\rightarrow(1,2)$ via elastic co-tunneling from the left lead (or direct tunneling through the almost closed right barrier) is opened up. The result is an increase in the occupation of the ground state configuration (1,2) (leading to the dark lines in differential transconductance). The sketches in Fig.\ \ref{sup_excited}b depict the relevant elastic channels of the transition $(1,1)\rightarrow(1,2)$. The deviation from the triangular shape of the back-action--induced region at large detuning in the lowest panel of Fig.\ 2e in the main paper is accordingly explained by an excited state in the right dot which strongly enhances the transition $(1,1)\rightarrow(1,2)$.

The interference pattern in Fig.\ \ref{sup_excited}a stems from transitions between the electronic ground states of the two dots mediated by the absorption of a phonon. Intradot transitions involving excited electronic states would cause characteristic phase shifts in the interference pattern, which we do not observe in our experiments. The two black lines in Fig.\ \ref{sup_excited} cross the interference pattern without disturbing it. We therefore conclude that the decay of electronic excited states is fast compared to the interdot transition rate.

\section{Temperature dependence}

\begin{figure}[ht]
 \includegraphics[width=\columnwidth]{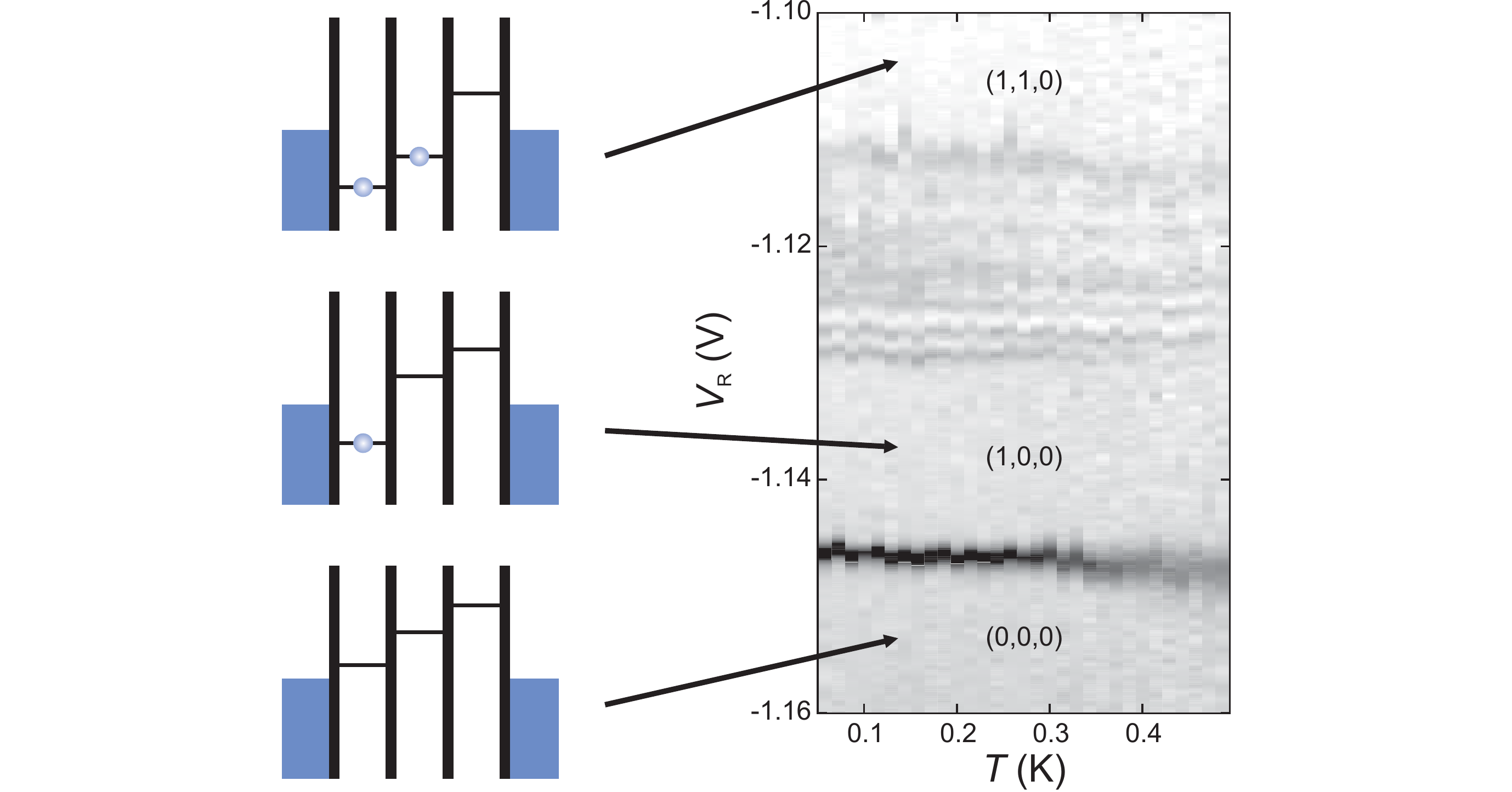}
\caption{\label{sup_temp}
{\bf Temperature dependence of the interference pattern:} Shown is a vertical slice through a stability diagram of the TQD as the ones in Fig.\ 3 of the main paper (but measured in a different cooldown) as a function of temperature for $\vqpc = \SI{-900}{\uV}$ and at constant $V_\text{L}=\SI{-1.272}{\V}$. On the left-hand side, the respective ground states are sketched. The dark charging line at the resonance $(0,0,0)\leftrightarrow(1,0,0)$ shows clear temperature induced broadening for $T>300\,$mK. The interference stripes above show no such broadening. They belong to the back-action triangle in the lower right corner in Fig.\ 3 of the main paper. Above a few clear interference stripes is the first charging line of the center dot obscured by back-action.}
\end{figure}

The temperature dependence of a slice through a stability diagram of the TQD (compare Fig.\ 3 of the main paper) is shown in Fig.\ \ref{sup_temp}. 
It includes the first charging line of the right dot (lower horizontal and dark line in Fig.\ \ref{sup_temp}) and a small number of clear interference stripes of the triangle in the lower right corner in Fig.\ 3 of the main paper. The washed out region above the clear interference stripes contains the first charging line of the center dot which is, however, smeared out by back-action. The width of the dark charging line  increases with temperature as can be clearly seen in Fig.\ \ref{sup_temp} for $T>300\,$mK. This effect is caused by electrons tunneling between the left dot and its lead and the temperature broadened Fermi-Dirac occupation of the electronic states in the lead. Considerably less temperature broadening is observed for the interference stripes. Such a weak temperature dependence is only possible if the interference pattern is based on interdot transitions which are not influenced by the temperature broadening of the Fermi edge in the leads. This observation clearly supports our interpretation of the observed back-action in terms of nonequilibrium phonons driving interdot transitions.

\section{Theoretical Modelling}

\begin{figure}[hb]
\begin{center}
\includegraphics[width=\textwidth]{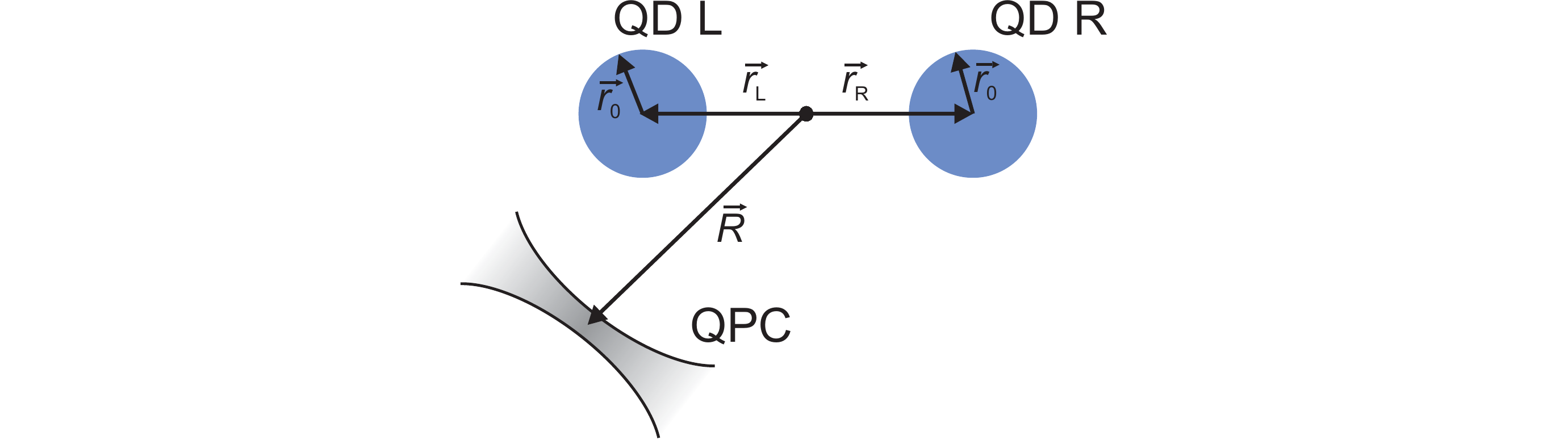}
\end{center}
 \caption{\textbf{Geometry of the double quantum dot with respect to the nearby QPC.}} \label{fig1}
\end{figure}

\subsection{Electron-phonon interaction}

As discussed in the main text, the back-action process of interest involves two charge configurations where an extra electron is in one of either two adjacent quantum dots (e.g.~in the DQD system described in Fig.~3, these are the charge states $(2,1)$ and $(1,2)$).  For simplicity, we present theory for the DQD system, and focus on the state of the extra electron.  The two relevant charge states are thus $\ket{L}$ (extra electron in left dot) or $\ket{R}$ (extra electron in right dot).
The Hamiltonian of the DQD takes the form
\begin{equation}
	H_{\rm dqd} = \ep\, \left(\ket{L}\bra{L}- \ket{R}\bra{R}\right) + t_c \,\left(\ket{L}\bra{R} + \ket{R}\bra{L}\right),
\end{equation}
where $\ep$ denotes the difference in electrostatic energies between the states $\ket{L}$, $\ket{R}$, and $t_c$
denotes the tunnel coupling between left and right dots.

Restricting attention to the subspace spanned by the states $\ket{L}$, $\ket{R}$, the Fourier transform of the DQD electron charge density operator at wavevector $\vec{q}$ takes the form:
\begin{equation}
\hat{\rho}_{\rm dqd}[\vec{q}] = \alpha_{\vq}\ket{L}\bra{L} + \beta_{\vq}\ket{R}\bra{R} + \gamma_{\vq} \ket{L}\bra{R} + \
\gamma_{-\vq}^* \ket{R}\bra{L}  \label{rho}
\end{equation}
where $\alpha_{\vq}=\bra{L} e^{i\vq\cdot\vr}\ket{L}$, $\beta_{\vq}=\bra{R} e^{i\vq\cdot\vr}\ket{R}$ and $\gamma_{\vq}=\bra{L} e^{i\vq\cdot\vr}\ket{R}$. We take the ground state wavefunctions of the left and right dots to be Gaussians of width $r_0$, centered at $\vr_L$ and $\vr_R$, respectively. The system geometry is shown in Fig.~\ref{fig1}. Similar to previous theoretical treatments of phonon emission by a DQD \cite{Brandes:1999,Gasser2009}, we neglect the coupling between the off-diagonal elements of the DQD density operator in Eq.\ (\ref{rho}) and phonons, as such terms only weakly modify the main effect due to the diagonal terms in $\hat{\rho}_{\vec{q}}$.

Using the expression for the charge density operator, we can now write the interaction between DQD electrons and phonons in the standard manner.  Similar to Ref.~\cite{Brandes:1999},
we focus on the interaction with acoustic phonons via the piezoelectric interaction.  Unlike Ref.~\cite{Brandes:1999}, we keep all details of the acoustic phonon spectrum of GaAs (i.e.\ anisotropic sound velocities, polarizations), as obtained from a standard elasticity-theory calculation \cite{Jasiukiewicz:1998}, and also include the effects of screening.  It is useful to write the electron-phonon interaction in terms of the eigenstates of $H_{\rm dqd}$ which we denote $\ket{g}$ (ground state) and $\ket{e}$ (excited state).  Keeping only those terms which can generate transitions between the DQD eigenstates, we obtain the interaction Hamiltonian $H_{\rm int}$ given in Eq.~1 of the main text.
The matrix element $\lambda_{\vq, \mu}$ appearing in this equation is the effective screened matrix element for the piezoelectric interaction of phonons in mode $\mu$ with the electron density of a single quantum dot;  the mode index
$\mu \in \{\text{L, ST, FT}\}$ refers to the longitudinal, slow transverse and fast transverse modes, respectively.  The matrix elements take the general form \cite{Kawamura92, Jasiukiewicz:1998},
\begin{equation}
	|\lambda_{\vq,\mu}|^2 = \frac{\hbar}{2\rho_M\mathcal{V}\omega_{\vq,\mu}} \,
	e^{-q^2\,r_0^2/2} \,\mathcal{M}^\text{PA}_{\vec{q},\mu} \, \mathcal{F}_{q_{\perp},\mu} \, \mathcal{S}_{q_{\parallel},\mu} \label{new}
\end{equation}
where $\rho_M$ is the mass density of GaAs and $\mathcal{V}$ is the appropriate crystal volume element. The phonon frequencies are $\omega_{\vq,\mu}=c_{\hat{q},\mu}\,|\vq|$, where the corresponding sound velocities $c_{\hat{q},\mu}$ depend on the wavevector direction and are calculated within elasticity theory \cite{Jasiukiewicz:1998}.  As discussed in the Methods section, the Gaussian factor $e^{-q^2\,r_0^2/2}$ is a cutoff coming from the finite size of each dot; we take $r_0$ small enough ($r_0 \simeq 2 $ nm) so that this cutoff plays no role.  We discuss each of the remaining factors in what follows.

$\mathcal{M}^\text{PA}_{\vec{q},\mu}$ is the bare (unscreened) piezoelectric coupling matrix element, and takes the standard form:
\begin{equation}
\mathcal{M}^\text{PA}_{\vec{q},\mu} = (2eh_{14})^2\,\left( \hat{q}_1\,\hat{q}_2 \,e^{\mu}_3 + \hat{q}_2\,\hat{q}_3 \,e^{\mu}_1 + \hat{q}_3\,\hat{q}_1 \,e^{\mu}_2 \right)^2
\end{equation}
where $h_{14} = 1.44 \, \mathrm{ V/nm}$ is the piezoelectric constant for GaAs \cite{PiezoParam}. Here, $\hat{q}_i=q_i/|\vq|$ is the normalized component of the phonon wavevector along the crystallographic axis $i\in\{x,y,z\}$, while $e^\mu_i[\vec q]$ is the projection of the given phonon mode's polarization vector onto the $i$-axis.

The form factor $\mathcal{F}_{q_{\perp},\mu}$ in Eq.~(\ref{new}) accounts for the suppression of the interaction of electrons with phonons having a large wavevector component normal to the plane of the 2DES.  Assuming a standard triangular form for the transverse confining potential of the 2DES, one obtains:
\begin{equation}
	\mathcal{F}_{q_{\perp},\mu} = \left| \int dz\,|\rho_0(z)|^2 e^{-i\,q_{\perp}z}\right|^2 , \quad
	\rho_0(z) = \theta(z) \, \sqrt{\frac{1}{2a}} \left(\frac{z}{a}\right) e^{-\frac{z}{2a}}
\end{equation}
where $\theta(z)$ is the unit step-function, $a = 3.5$ nm is the 2DES thickness and $q_{\perp}$ is the component of the wavevector perpendicular to the 2DES plane. Here $\rho_0(z)$ is the transverse wavefunction of a 2DES electron.

Finally, the factor $\mathcal{S}_{q_{\parallel},\mu} $ describes the effect of screening in the plane of the 2DES.
Using a standard RPA approach which accounts for the two-dimensional nature of 2DES electrons \cite{Price:1981}, one obtains:
\begin{equation}
	\mathcal{S}_{q_{\parallel},\lambda} = \left( \frac{r_s\,q_{\parallel}}{H + r_s\,q_{\parallel}}\right)^2 , \quad
	H = \int \! dz' \, |\rho_0(z')|^2 \int \! dz\, |\rho_0(z)|^2 e^{-q_{\parallel}|z-z'|}
\end{equation}
where the effective screening radius $r_s = 5$ nm is equal to half the Bohr radius of GaAs, and $q_{\parallel}$ is the in-plane component of the phonon wavevector.
As expected, the screening factor suppresses the contribution from long-wavelength phonons, and thus suppresses the back-action stripe pattern at small values of the energy detuning $\Delta$.

\subsection{Quantum point contact charge noise spectrum}

As discussed in the main text and Methods section, the fluctuating electronic charge density associated with the QPC locally generates non-equilibrium acoustic phonons.  The strength of these charge fluctuations are described by the quantum noise spectrum $S_{QQ}[\omega]$ of the QPC charge operator $\hat{Q}$:
\begin{equation}
	S_{QQ}[\omega] \equiv \int_{-\infty}^{\infty} dt e^{i \omega t} \langle \hat{Q}(t) \hat{Q}(0) \rangle
\end{equation}
This charge-fluctuation spectrum can be calculated using a standard scattering-theory approach to mesoscopic transport
 \cite{Pedersen1998,Pilgram:2002,Young2010}.  The relevant, negative frequency part of the spectrum (which describes the emission of energy by the QPC) takes the following form at low temperature :
  \begin{equation}
S_{QQ}[-|\omega|]  = \frac{\hbar}{8\pi}\,\left(\frac{\Delta \TT}{\Delta U}\right)^2\frac{1}{\TT\,(1-\TT)} \,\left(eV_{\rm qpc}-\hbar|\omega|\right)\, \Theta(eV_{\rm qpc}-\hbar|\omega|) . \label{SQQ}
\end{equation}
Here, $\TT$ is the transmission of the QPC, $\Delta \TT$ ($\Delta U$) is the change in QPC transmission (potential) resulting from changing the DQD charge state from $\ket{L}$ to $\ket{R}$, and $V_{\rm qpc}$ is the QPC bias voltage.  Note that the magnitude of these charge fluctuations is set by the sensitivity of the QPC to the DQD charge state-- this is a direct consequence of these charge fluctuations being the fundamental Heisenberg back-action of measurement with a QPC
\cite{Young2010}.

\subsection{Master equation approach}

As discussed in the Methods, the theory calculation involves two initial steps:
\begin{enumerate}
	\item
		We first describe the generation of non-equilibrium ``hot" acoustic phonons by the QPC charge fluctuations.
		We do this by calculating
		the Keldysh Green functions of the acoustic phonons to first order in the electron-phonon coupling to the QPC.
		This coupling Hamiltonian takes the form:
		\begin{equation}
			H_{\rm int, qpc} = \sum_{\mu, \vec{q}}   \lambda_{\mu, \vec{q}} \, e^{i \vec{q} \cdot \vec{R} }  \, \hat{Q}  \left(
			\hat{a}^{\phantom{\dagger}}_{\mu, \vec{q} } + \hat{a}^{\dagger}_{\mu, -\vec{q} } \right)
		\end{equation}
		The ``heating" correction to the phonon Green functions can be expressed in terms of the charge noise spectrum
		$S_{QQ}[\omega]$ given above.  Here $\vec{R}$ denotes the position of the QPC with respect to the midpoint
		between the two quantum dots (see Fig.~1).  Similar to our treatment of the dots, the spatial extent $r_0$
		of the QPC
		charge distribution serves as a high-energy cutoff in the above interaction; we take this scale to be small
		enough that it plays no significant role (i.e.~the QPC voltage instead provides the relevant cutoff).
	\item
		We next calculate Golden rule rates $\Gamma_{\ua}, \Gamma_{\da}$
		describing transitions between the DQD states $\ket{g}$ and $\ket{e}$
		via the DQD-phonon interaction given in Eq.~1 of the main text; this is done using the ``heated" phonon
		Green functions computed above.
\end{enumerate}

Finally, we incorporate the rates $\Gamma_{\ua}, \Gamma_{\da}$ into a master equation
describing the probabilities of the states
$\ket{g}$, $\ket{e}$ (which have a total of 3 DQD electrons), as well as the two-electron state $(1,1)$ (denoted $\ket{ 2 }$)
and the four-electron state $(2,2)$  (denoted $\ket{ 4 }$).  This is similar
to the approach outlined in Ref.~\cite{Gasser2009}.  This master equation describes the electrostatic blocking mechanism depicted in Fig. 4 of the main text.  In addition to the phonon-assisted rates, the master equation also involves  rates describing incoherent lead tunneling to and from the DQD.  A slow incoherent rate $\Gamma_{\mathrm{slow}}$
(involving lead tunneling from the right) describes transitions from state $\ket{2}$ to $\ket{g}$, and a fast incoherent rate
$\Gamma_{\mathrm{fast}}$ (involving tunneling to the left lead) describes transitions from state $\ket{e}$ to $\ket{1}$.
These rates (which are set by the tunnel coupling to the leads) also determine the incoherent rates describing transitions
from $\ket{e}, \ket{g}$ to the $(2,2)$ state.

We are interested in the experimentally relevant limit where $\Gamma_{\mathrm{fast}} \gg \Gamma_{\mathrm{slow}}$, corresponding to the conditions underlying the  blocking mechanism.  In this limit, and for gate voltages far from the charging-lines for the $(2,2)$ state, the stationary probability $P_g$
to be in the DQD ground state $|g\rangle$ takes the simple form:
\begin{equation}
	P_g
	 \rightarrow \frac{\Gamma_{\mathrm{slow}}}{\Gamma_{\ua} + \Gamma_{\mathrm{slow}}} .
	 \label{eq:PG}
\end{equation}
Importantly, the fast rate $\Gamma_{\mathrm{fast}}$ does not enter the limiting equations when $\Gamma_{\mathrm{fast}} \gg \Gamma_{\mathrm{slow}}$. In addition, it is straightforward to show that the relaxation rate $\Gamma_{\da}$ connecting states $\ket{e}$ and $\ket{g}$ does not affect this result so long as $\Gamma_{\da} \ll \Gamma_{\mathrm{fast}}$. The magnitude of the ground state occupation therefore depends only on the relative magnitude of the slow refilling rates and the coherent excitation rate.

Finally, we can use the master equation to calculate the derivative $\dif n_{\rm eff} / \dif V_{L}$, where $V_{L}$
is the gate voltage used in the experiment to extract the differential transconductance (see Fig.~3 of the main text), and
\begin{equation}
	\bar{n}_{\rm eff} = \langle n_R \rangle + \varepsilon \langle n_L \rangle
\end{equation}
is the effective charge sensed by the QPC.  This quantity is proportional to the experimentally measured differential transconductance
$\dif I_{\rm qpc} / \dif V_{L}$, thus allowing a comparison between theory and experiment.  The parameter
$\varepsilon = 0.4$ is determined experimentally from the QPC's relative sensitivity to charge addition to the L dot versus the R dot.

\subsection{Numerical and experimental parameters}

The theoretical plots in the main text take the distance between QDs in Fig.~\ref{fig1} to be $d=|\vr_L - \vr_R|\approx 235\,$nm;
this is in reasonable agreement with estimates made from SEM images of the device, and also yields a spacing between back-action--excitation lines in the stability diagram that match experiment.  Based on estimates from device images, we take the DQD-QPC separation to be $R\approx 500\,$nm.  Electrostatic energies used in the theory are obtained from experimentally measured charging diagrams.  We find charging energies of the left and right dots to be $E_{C,L}=2.9$ meV and $E_{C,R}=2.7$ meV , respectively, while the interdot charging energy is $E_{Ci}=0.5\,$meV. We also use a
value of interdot tunnel coupling $t_c \approx 7\mu\,$eV
(being half of the energy splitting between
$\ket{e}$ and $\ket{g}$ for $\varepsilon = 0$)
that is extracted from measured stability diagrams.  Finally, as
already discussed, the theoretical calculations take the spatial extent of dot and QPC charge distributions to be small enough that they play no role ($r_0 = 2$ nm).

The DQD experiments described in this article employ a small QPC transmission $\TT \approx 0.0028$ in order to avoid back-action due to shot noise and heating effects due to large QPC powers. The change in transmission associated with a change in the DQD charge state is measured to be $\Delta \TT \approx 0.00043$. Via a simple calculation using a screened Coulomb potential, one can show that moving the excess electron from the left dot to the right results in a potential change of $\Delta U \approx 25\,\mu$V at the position of the QPC, which is consistent with experimental estimates. The QPC biases employed in our measurements are on the order of $V_{\rm QPC} \sim 1$ mV.
This procedure is used as a convenient way to determine the slope of the QPC's transmission curve as a function of local potential changes.

For the incoherent rates $\Gamma_{\mathrm{fast}}$ and $\Gamma_{\mathrm{slow}}$ connecting the undriven DQD to the leads, and entering our master equation calculation, only rough estimates are available from our experiments.
For our numerics, we take $\Gamma_{\mathrm{slow}} = 10$ kHz, and set $\Gamma_{\mathrm{fast}} = 1$ GHz such that $\Gamma_{\mathrm{fast}} \gg \Gamma_{\mathrm{slow}}$.  We stress that in this limit, the specific value of $\Gamma_{\mathrm{fast}}$ plays no role.  The value of the slow rate $\Gamma_{\mathrm{slow}}$ only serves to determine the overall magnitude of the back-action--induced probability oscillations, cf.~Eq.~(\ref{eq:PG}).


\end{document}